# IEEE Copyright notice

# Investigating the refractive index sensitivity of U-bent fiber optic sensors using ray optics

Christina Grace Danny, M Danny Raj, VVR Sai

*Abstract* Geometrically modified fiber optic sensors (FOS), particularly U-bent FOS, have gained significant attention due to their remarkably high refractive index (RI) and evanescent wave absorbance (EWA) sensitivity, as well as their ergonomic design and ease in handling. In this study, we present a theoretical model for the U-bent FOS probes, to predict the sensor behavior by numerically simulating the light propagation in an equivalent 2D semi-circular ring using ray tracing approach. In addition to the effects due to the modification of geometry, this study presents a thorough investigation of the influence of the bend-induced material deformation on the nature of light propagation and refractive losses. We introduce 'bend ratio' (ratio of bend radius to fiber core radius), to explain the influence of geometry modification and the bend-induced inhomogeneity in RI (BIRI) of the fiber core on RI sensitivity. The bend ratio of bent plastic optical fiber sensors falls under one of the four bending regimes namely, gentle, geometric, saturation and plastic, for which the bend ratios are less than ~35, ~25, 17 and 7 respectively. The results also show that for bend ratios less than 7, BIRI inhomogeneity is responsible for the high RI sensitivity observed with U-bent probes as opposed to the simple geometric modification. This study also indicates the existence of an optimum bend ratio (for a given value of RI of the surrounding medium) where RI sensitivity is maximum. These findings were validated with previously reported experimental results.

*Index Terms*— Optical fiber sensors, Geometric optics, U-bent FOS

## I. INTRODUCTION

In the recent years, there has been an unparalleled motivation towards realizing biochemical sensors for healthcare, food and environmental monitoring [1], [2]. Optical sensing techniques have been widely explored for this purpose due to the myriad advantages including portability, sensitivity and high throughput [3], [4]. Fiber optic sensors (FOS) and geometry-modified FOS in particular have gained attention as a potential candidate for cost effective and quantitative biochemical sensing. Geometrical modifications are introduced in a FOS mainly to improve the evanescent wave-based interaction of light with the sample solution/ surrounding medium. There are several probe configurations that have been reported, including tapered, etched, D-shaped and U-bent. Amongst them, the U-bent probes have been reported to have excellent refractive index (RI) sensitivity as well as evanescent wave-based absorbance (EWA), as high as 10-fold in comparison to straight decladded fiber probes. One unique advantage of U-bent optical fiber over the above geometries is its ergonomic design. The FOS probes are compact, with its small and robust sensing region on one side and its two distal ends on the other side. They have a dip-type probe configuration with a small form factor of the sensing region that is optimum for efficient interaction with small volumes of samples. The optical coupling is also simple, with light source and photodetector at each of the distal ends.

Over the past three decades, a large number of studies utilizing U-bent FOS probes have been reported for a variety of applications including refractive index based sensing [5], optical absorbance spectroscopy [6], localized surface plasmon resonance (LSPR) based sensing [7], [8], pH [9], temperature [10], humidity [11], chemical and biosensing [12], [13]. A significant number of those studies deal with the optimum geometry for the U-bent probes since it plays a major role in realizing a highly sensitive optical transducer for various applications [5], [8], [10], [14]–[16]. B D Gupta et al have carried out experimental as well as theoretical investigations to understand the influence of bend radius, fiber core radius and numerical aperture on the RI and EWA sensitivity [6], [17]. In addition, Sai and co-workers [5], [8], [12] have experimentally demonstrated that the ratio of bend diameter to the core diameter is a possible measure to determine optimum geometric conditions. They also reported that U-bent FOS with bend diameter less than 5 times the fiber core diameter is highly sensitive to RI changes.

Several theoretical models have been developed to understand light propagation and loss in a bent optical fiber. Light attenuation in a bent fiber has mostly been studied by solving Maxwell's equations with appropriate boundary conditions (for fiber core <50 µm) [18]; or by using Fresnel's coefficients integrated over a range of incidence angles (for fiber core >50 µm) [10]. In most cases, the literature is limited to understanding the bend losses in a cladded optical fiber meant for communication purposes, where weakly guiding approximation is valid [19], [20]. However, in the field of FOS, fibers are typically decladded for effective interaction with the medium, making the decladded region – strongly guiding, demanding newer models [21]. These models could be

Christina Grace Danny was with the Department of Applied Mechanics, Indian Institute of Technology Madras, Chennai 600036, India and is currently with the Department of Electronics and Instrumentation, Ramaiah Institute of Technology, Bengaluru 560094, India. (e-mail: christina@msrit.edu).

M Danny Raj was with the Department of Chemical Engineering, Indian Institute of Technology Madras, Chennai 600036, India and is currently with the Department of Chemical Engineering, Indian Institute of Science, Bengaluru 560094, India (e-mail: dannyrajmasila@gmail.com, dannym@iisc.ac.in).

V V Raghavendra Sai is with the Department of Applied Mechanics, Indian Institute of Technology Madras, Chennai 600036, India (e-mail: vvrsai@iitm.ac.in).

developed either using wave optics or ray optics approaches. Due to the computational limitations of the wave optics model, ray optics models have been preferred for FOS utilizing multimode fibers of large diameters [22]. Still, the ray optics models developed thus far for acutely bent FOS offer only a limited understanding of light propagation and are yet to provide a comprehensive reasoning for the high RI and EWA sensitivities reported [16], [23]–[25].

In this study, light propagation in a decladded U-bent FOS placed in an aqueous medium of uniform RI is investigated using ray optics. The work attempts to address the contribution of refractive losses to the RI sensitivity by the probe geometry as well as the bend induced material deformation, the evaluation of sensitivity and linearity over a broad range of RI values of the surrounding medium and more importantly obtain estimation of optimum fiber probe parameters for the maximum sensitivity. In this article, we begin with the description of the theoretical model developed based on the ray optics approach and the relevant equations and assumptions made. Then, the numerical results are compared with that of other analytical approaches. Subsequently, the results obtained from the model are explained considering refractive losses (i) due to geometry alone and subsequently, (ii) due to bend induced inhomogeneity in RI of core in combination with geometric effects. Finally, the estimated sensitivity is compared with experimental results.

## II. RAY OPTICS BASED MODEL FOR U-BENT FOS

A U-bent FOS is mostly a multimode optical fiber (of ~ 20 cm length), decladded in the middle and bent to a desired bend radius. The ends of the U-bent FOS are connected to a pair of LED and photodetector [5]. When, the decladded bent portion is placed in a reference medium of known RI value, a certain portion of the light that is guided by the straight region of the optical fiber is lost at the bend region, which is known as refractive loss. This refractive loss is proportional to the RI of the surrounding medium. The refractive index of an unknown medium ($n_{med}$) can be estimated by quantifying the change in the losses with respect to that of the reference medium. The U-bent FOS with very low bend radii (<10× of the fiber core radius) have a very high optical loss and display remarkable sensitivity to the surrounding medium.

In general, bending an optical fiber is known to introduce refractive losses. The analytical models proposed for estimating the bending losses are mainly limited to single and few mode fibers [26]. However, in the case of U-bent multimode FOS, it is difficult to formulate a purely analytical framework due to the abrupt changes in the direction of light propagation in the highly deformed bend region, and the inhomogeneity in both material and the geometrically defined numerical aperture. Thus, we utilize ray optics approach to gain an insight into the light propagation characteristics for U-bent FOS. The proposed model aims to capture the trajectory of rays coupled into the FOS at various incident angles.

The U-bent FOS is modelled as a 2D semi-circular ring with an RI of $n_{co}$, placed in an aqueous medium of varying RI, $n_{med}$ (Fig. 1) [27]. The light rays enter from the left side of the semi-circular ring and exit at the right side of the semi-circular ring. The propagation direction is along the optical axis of the fiber. The input parameters to the model could be categorized into a) **material parameters** – RI of core ($n_{co}$), cladding ($n_{cl}$), medium ($n_{med}$); the elasto-optic coefficients of the fiber optic core (Pockel's coefficients - $P_{ij}$); and the Poisson's ratio ($\gamma$), b) **geometric parameters** – fiber core radius ($\rho$) and bend radius (R), and c) **simulation parameters** –number of point sources distributed over the cross-section ($N_p$), number of angles considered at each point source ($N_a$), and the originating point and its irradiance angle ($\theta$) for each ray. These inputs are determined as detailed in this section (II.A). The power loss at each core-medium interface is given by Fresnel's coefficients. Using the model, the trajectories of the rays coupled into the FOS are simulated and the **output parameters** including guided and lost power, sensitivity, linearity and width of propagation are estimated.

### A. Optical source and coupling

At the entry region of the U-bent fiber, a large number of point sources distributed over the cross-section of the fiber ($p_1$, $p_2$, $p_3$…), and several rays originating from each point source at various angles ($\theta_1$, $\theta_2$, $\theta_3$…) are considered, as shown in Fig. 1(thus, the model is independent of the length of straight fiber before bending). The fiber core diameter (>200 μm) and numerical aperture (NA >0.3) of the fiber considered here give rise to a very large V number, resulting in a highly multimode light propagation. The light rays that could be coupled to the optical fiber for propagation by total internal reflection, spans from $\theta_c$ to 180−$\theta_c$, where critical angle, $\theta_c$ is given by $\sin^{-1}(n_{cl}/n_{co})$. The rays subtend an angle $\theta$ to the transverse cross-section of the waveguide and subtends (90−$\theta$) to the propagation direction. The model considers only the propagation of the meridional rays that travel in the plane of the fiber bending. Meridional rays propagating in all the other planes and all skew rays are not considered in the current model.

The light is assumed to be coupled from the source through a microscopic objective on to the optical fiber end face at the axial point (such that, power of the rays traveling close to the critical angle is lower than that of the rays traveling parallel to the fiber core) [17]. The incident power profile P($\theta$), is modeled using (1). (Refer Supporting information Sec. A for Fig. S1).

$$P(\theta) = \frac{n_1^2 \sin\theta \cos\theta}{(1 - n_1^2 \cos^2\theta)^2} \quad (1)$$

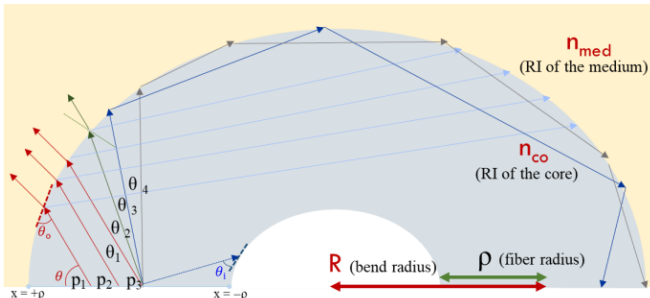
Fig. 1: Ray optics model for evaluation of refractive losses in a U-bent fiber optic sensor

## B. Tracing of rays in a U-bent FOS

A pictorial representation of ray tracing is given in Fig. 1. The ray that subtends an angle θ to the transverse cross-section of the waveguide, propagates in the same direction till it intersects the core-medium interface. The point of intersection of a ray with the core-medium interface at the bent segment was geometrically estimated by considering the ray segment as a straight-line intersecting the semicircular outer or inner core-clad interface. Thus, the rays subtend $\theta_o$ or $\theta_i$ (angles of incidence) with respect to the tangent at the outer and inner interfaces respectively.

After an intersection, the ray reflects or refracts at the core-medium interface as given by Snell's law (2). The conditions for direction of propagation and total internal reflection were obtained from (2) and (3) respectively.

$$n_{co} \sin \theta_i = n_{med} \sin \theta_r \quad (2)$$

$$\sin \theta_c = \left(\frac{n_{med}}{n_{co}}\right) \quad (3)$$

where $\theta_i$, $\theta_r$ and $\theta_c$ are incident, refracted and critical angles respectively. $n_{co}$ and $n_{med}$ are the RI of the optical fiber core and the surrounding medium respectively. Fresnel's coefficients were used in this ray tracing algorithm to precisely account for the refraction and reflection of the rays at the core-medium interface. (Refer Supporting information Sec. B for details.)

## C. Estimation of U-bent FOS sensitivity

Refractive loss and sensitivity were estimated as per (4) and (5) respectively.

$$Normalized\ Refractive\ Loss\ (dB)_{med} = 10 \log \left(\frac{P_{ref}}{P_{med}}\right) \quad (4)$$

$$Sensitivity = \frac{\Delta Loss(dB)}{\Delta RI} \quad (5)$$

where $P_{ref}$ is the power measured in a reference medium, usually water ($n_{ref}$=1.33) and $P_{med}$ is the power received at the other end of the optical fiber for the respective RI of the surrounding medium. (Experimentally $P_{in}$ – the power coupled into the optical fiber is not measurable, hence $P_{ref}$ is used.)

The sensitivity of U-bent FOS is estimated in two stages. In the first stage, the sensitivity due to the U-bent geometry was evaluated and is discussed in III.A. Further, in the second stage, the thermal and mechanical stresses developed in the fiber core material due to the bending process were taken into consideration. These stresses result in alteration of material parameters, especially the RI of the fiber core, significantly affecting the light propagation and thus the sensitivity of the FOS (as discussed in III.B). This effect is incorporated as **B**end **I**nduced **RI** inhomogeneity in the fiber optic core. The extent of variation in the core RI due to the stresses was estimated using fiber core material properties including elasto-optic coefficients and Poisson's ratio.

## D. Evaluation of the model
### 1) Convergence of model parameters

The accuracy of a ray tracing model is highly influenced by the number of rays considered in the simulation, which is a product of $N_a$ and $N_p$. Ideally, the number of rays (based on V number) was calculated as per (6) [28].

$$N = 0.5 \times \left(\frac{\pi \times Fiber\ Diameter\ \times Numerical\ aperture}{\lambda}\right)^2 \quad (6)$$

However, simulations involving such a large number of rays (~$10^5$) are computationally intensive, making the model unsuitable for optimal parameter identification studies. Hence, a minimum number of rays with ($N_a$ = 100 and $N_p$ =100), for which the output parameter converged, was used in all subsequent simulations (see Supporting information Sec. C).

### 2) Analytical and numerical estimation of local NA

The analytical expression for local numerical aperture (NA), that is characteristic of the light guidance capacity of a geometrically local point in a bent optical fiber can be determined using (7) given by Remouche et al [10].

$$\mathrm{NA} = \sqrt{\left[n_{co}^2 - n_{med}^2 \left[\frac{R+\rho}{R+x}\right]^2\right]} \quad (7)$$

By applying (7), the local NA of a plastic optical fiber of 0.5 mm core diameter was estimated for various bend radii (R) and is represented as solid lines in Fig. 2. The local NA of a bent fiber at sufficiently low bend radii reduces to zero towards the inner curvature of the bend and subsequently only a part of the fiber core has a positive non-zero NA (Fig. 2). This signifies that light propagates only in a small part of the optical core in case of an acutely bent optical fiber.

Using the U-bent FOS model developed here, simulations were run for the above-mentioned conditions. For each ray that was traced, the minimum radial distance (x) of the traced ray segment from the origin was estimated and the width of propagation was defined as (R+ρ-x). Then, the minimum width of propagation for all guided rays was estimated and represented as dashed straight lines in Fig. 2. Interestingly, the distance at which the NA becomes 0 as estimated by the analytical equation (7) significantly matched with the width of propagation as evaluated from the numerical 2D ray tracing algorithm for (i) different bend radii and (ii) different medium RI (refer Supporting information Sec. D). This implies that the ray tracing algorithm which is commonly used to determine only optical power loss, could also be used to estimate parameters such as width of propagation, that gives a better understanding of the light propagation characteristics in a bent optical fiber.

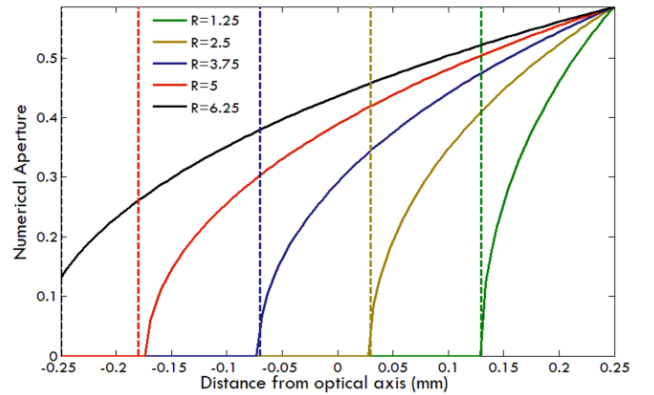

Fig. 2: Simulation results obtained for a plastic optical fiber of 0.5 mm fiber core diameter. The NA across the various bend radii for $n_{med}$ of 1.37 was calculated using analytical equation and compared with the width of propagation estimated using ray optics model

### III. RESULTS AND DISCUSSIONS

#### A. Refractive loss due to geometry

In a straight optical fiber, the light rays maintain the same angle of incidence throughout its length. In a significantly bent optical fiber, the light rays strike the outer curved surface of the bend at a lower angle (with respect to the normal drawn at the bent surface). Further, the critical angle at the decladded bent region itself is expected to reduce due to the presence of a medium whose RI is much lower than that of the cladding. When the altered angle of incidence at the bend region is lower than the critical angle, the corresponding ray is lost due to refraction. This is known as bending loss of the optical power coupled into the fiber probe. In addition, it should be noted that all the rays that are lost in the bend region are lost at the outer interface.

Typically, a POF made of PMMA has core and cladding RI as 1.49 and 1.41 respectively. In accordance with Snell's law, the light rays with an incident angle of 71° to 109° continue to propagate further along the optical fiber (based on core and cladding RI). When this POF is bent, decladded and placed in aqueous solutions with $n_{med1}$=1.33 and $n_{med2}$=1.37, the critical angles for the cladded region, and the decladded region surrounded by medium 1 and 2 are 71, 63.2 and 66.84° respectively. In the bent region, the rays with incident angles lower than 63.2° are lost (refracted) in both the mediums. Similarly, the rays that have incident angles greater than 66.84° are guided in both the media Subsequently, the rays having incident angles lesser than 66.84°, and greater than 63.2° contribute to the medium-sensitive refractive loss. The critical angle at the decladded region and the modified propagation angle at the bend region depends on the RI of the medium and the probe bend radius respectively. Thus, refractive loss in a U-bent FOS is sensitive to the surrounding medium, due to geometric effects as elaborated further in this section.

#### 1) Normalized refractive Loss (dB)

The normalized refractive loss (dB) for 0.5 mm POF U-bent FOS for various bend radii was estimated using (4) for the aqueous media with RI values in the range of 1.33-1.37 with 1.33 as reference, as shown in Fig. 3. The refractive loss as simulated using the U-bend FOS model increases as the bend radius decreases. The trend described by the simulation results matches with the experimental observations made by Gowri and Sai [5], although the exact values of the bend radii differ.

#### 2) RI Sensitivity

One of the primary objectives of the U-bend FOS model is to identify the geometric parameters of the U-bent FOS that result in maximum RI sensitivity. Our previously reported experimental investigations [5] reveal that the 'bend ratio' - a dimensionless quantity defined as the ratio of bend radius to fiber radius, is a critical parameter to optimize for maximum possible RI sensitivity [5]. This result is in congruence with the simulation results using the model presented here, where the refractive losses are dependent only on the ratio of bend radius to fiber radius (verified by simulations). The RI sensitivity of each of the probe geometries obtained through the simulations (calculated as the slope of the responses at RI = 1.37) is shown in Fig. 4. For bend ratios higher than ~35, bending has a negligible influence on the sensitivity of probe. Hence, we term this region as **gentle bend**. The RI sensitivity increased steeply for bend ratio lower than ~25 ($R_{geometric}/\rho$) till it reaches a bend ratio of ~17 ($R_{saturation}/\rho$). This region where geometric effects are prominent is termed as **geometric bend**. It is important to note that the sensitivity plateaus for the bend ratio below ~17, where geometric effects do not contribute to any further enhancement in the sensitivity.

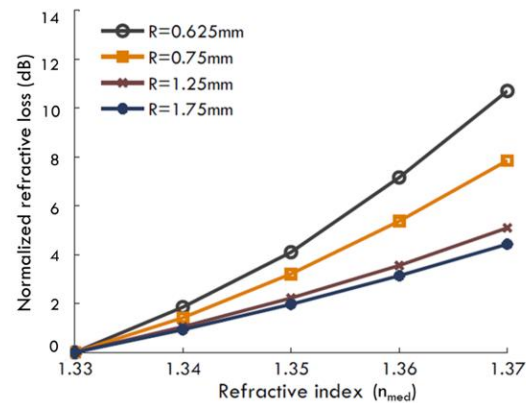

Fig. 3: U-bent POF sensor response of normalized refractive loss (dB) versus refractive index of the surrounding medium for various bend radii (POF diameter, 0.5 mm) as predicted by the model (simulations)

The existence of geometric bending regime demarcated by the transition radii, $R_{geometric}$ and $R_{saturation}$, where the geometric effects are prominent was verified by other means including estimation of (i) modified angle at the outer interface, $\theta_o$, and (ii) local numerical aperture. These results substantiate the maximum losses due to geometry between the saturation and the geometric bend radii (see Supporting information Sec. E).

The simulation results were compared with those obtained from similar 2D [23] and 3D [16] ray tracing models reported earlier and found to be qualitatively similar. However, the experimental observations show a significant or drastic increase in the sensitivity of U-bent FOS below bend ratio of 7 [5]. The U-bend FOS model incorporating only geometric effects is unable to reproduce or predict this behavior.

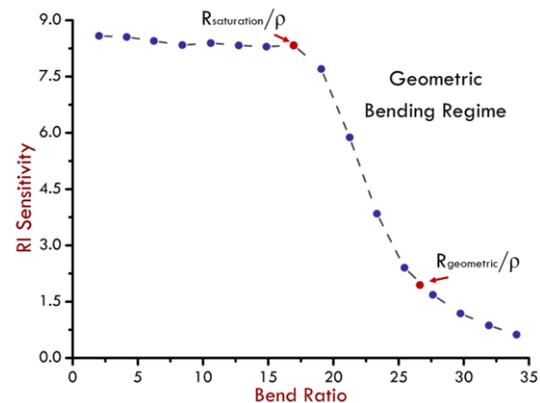

Fig. 4: RI sensitivity of U-bent POF for various bend ratios, considering $n_{med}$ =1.37. (RI Sensitivity is estimated as change in normalized refractive loss (dB) for a change in the RI of medium with respect to water as reference.)

### B. Refractive loss due to BIRI Inhomogeneity

*1) Photoelasticity and non-linear strain effects*

Bare PMMA is an optically isotropic material. However, fabrication of U-bent probes involves application of opposing force gradients about the cross-section of a POF (tensile to compressive forces from outer to inner curvatures respectively) along its optical axis, which induces shear strain and bending moments in the U-bent region. A subsequent thermal treatment over a short 10 min duration (to heat it to its glass transition temperature) results in an anisotropic core with residual stresses [5]. As a consequence, the nature of strain introduced in the bend region is highly tensile at its outer curvature (ρ+x) while it is highly compressive at the inner curvature (ρ-x) (with an assumption of zero strain along the optical axis, x = 0). An alteration in the light propagation is anticipated due to the strain developed in the fiber by the applied forces, that could be due to (i) Change in dimensions of the fiber, which can be estimated using mechanical properties such as elasticity coefficient and Poisson's ratio [29]. However, these changes are negligible as observed from the optical microscopic images. (ii) Alteration in the otherwise homogeneous RI profile of the fiber cross-section across the bend region due to photoelasticity, termed as bend-induced RI (BIRI) inhomogeneity. This effect could be predominant, as discussed below.

The alteration in light propagation due to strain can be described effectively with the help of phase shift. The equation pertaining to phase shift and applied strain (8), and the relationship between phase shift and refractive index of optical fiber core (9) are given below.

$$\Delta\varphi = \frac{d\varphi}{d\epsilon}\epsilon + \frac{1}{2}\left(\frac{d^2\varphi}{d\epsilon^2}\right)\epsilon^2 + \frac{1}{6}\left(\frac{d^3\varphi}{d\epsilon^3}\right)\epsilon^3 \qquad (8)$$

$$\Delta\varphi = \frac{2\pi}{\lambda}L(\Delta n + \epsilon n) \qquad (9)$$

where φ is the phase retardation, ε is strain and n is the refractive index of the optical fiber core, L is the length of the fiber, λ is the wavelength of the light coupled in the fiber.

PMMA is known to undergo elastic deformation for strains within ~2-3% and subsequently a plastic deformation up to ~15% strain. In the elastic range involving only small deformations, the stress or strain induced changes in RI (Δn) are linear and of the first order (only the first term in (8)). Incorporating bending stress in the above equation results in Neumann-Maxwell stress equations (10) and (11) [30], [31].

$$\Delta n_{ij} = P_{ijkl}\epsilon_{kl} \qquad (10)$$

$$n_x = n_{co}\left[1 - \left[\frac{n_{co}^2}{2}\right][P_{12} - \nu(P_{11} + P_{12})]\left(\frac{x}{R}\right)\right] \qquad (11)$$

where Δn is the RI tensor, P is Pockels tensor, ε is the strain tensor components, $n_x$ is the effective RI at the distance 'x' from the center of fiber, $n_{co}$ is RI of the fiber optic core under no stress, $P_{11}$ and $P_{12}$ are Pockels or elasto optic coefficients, ν is Poisson's ratio, R is the bend radius. Using (11), the change in the refractive index of the optical fiber core could be estimated for a wide range of bending profiles [31], [32], [33]. Here, two independent photoelastic coefficients (Pockels coefficients - $P_{11}$ and $P_{12}$) are sufficient to evaluate phase shift (Δφ).

Bending an optical fiber to a U-shape results in a material deformation that goes beyond the elastic to the plastic regime. In this regime, the higher order strain terms in (8) become more critical in evaluating the photoplastic behavior accurately [29]. In this case, additional photoelastic coefficients are required to evaluate Δφ, which have not been identified so far. The contribution of the higher order terms is necessary to explain the behavior of the optical fiber for large strains that result in failure and could be ignored for lower strains [29]. In addition to the above equation, accurate estimation of RI profile at a given cross-section also involves consideration of (i) the thermal treatment during the process of fabrication of the U-bent optical fiber probes such as temperature used for thermosetting or partial annealing, and (ii) wavelength dependency of the photoelastic constants [34].

Due to the insufficiency of experimental studies on evaluation of the above parameters for PMMA (for the case of acute bending) and the complexity involved in designing such characterization studies for bent optical fibers, this study uses a simplified photoelastic model considering only first order photoelastic terms, which may impose certain limitations on the accuracy of the model [35]. Also, the influence of thermal treatment on residual stresses and birefringence are not considered since the minimization of the residual stresses typically requires annealing over a long duration (of 8 h to 4 days), while the U-bent POF probes were subjected to only brief thermal treatment during the fabrication process [5].

*2) BIRI inhomogeneity*

In this section, we assume that the material deformation is only due to a bending force. Thus, bending moment (a function of bend radius) and the material parameters (such as Pockels or elasto-optic coefficients, Poisson's ratio) determine the change in the core RI. The BIRI inhomogeneity for a 2D geometric model of the bent fiber region (where only the light propagation in a single plane) could be estimated using (11) [21], [31], [31],[36].

The material properties including Poisson's ratio and Pockels coefficients for a PMMA optical fiber are taken from Sánchez et al [31] as ν=0.37 (the mean of the range 0.35 – 0.4 mentioned in literature) and $P_{11}$ = 0.121, $P_{12}$ = 0.270. The effective core RI for bent optical fibers made of PMMA is thus given by (12).

$$n_x = 1.49\left(1 - 0.1391 \times \left[\frac{x}{R}\right]\right) \qquad (12)$$

The variation in the RI of the optical fiber core of 0.5 mm POF with 1 mm bend radius is shown in Fig. 5. There is a significant reduction in the core RI at the outer curvature (from 1.49 to 1.44) and similarly a significant increase in the RI at the inner curvature (from 1.49 to 1.54). A variation of 0.1 RI units from the inner to the outer interfaces for bend ratios as small as 4 affects the light propagation and results in significant refractive losses as described in the subsequent sections.

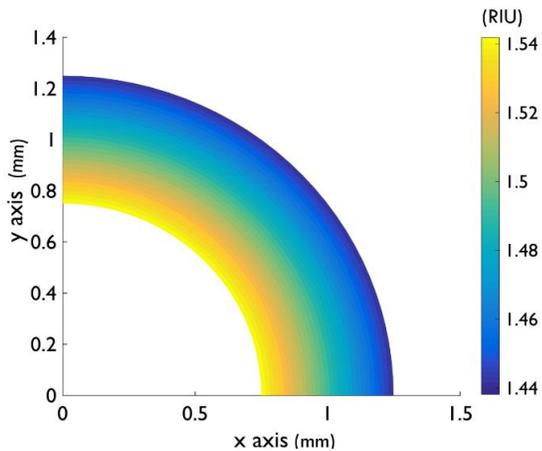

Fig. 5: Variation in core refractive index due to material deformation caused by bending

To incorporate BIRI inhomogeneity, the ray tracing algorithm was modified to accommodate the gradient RI profile of the optical fiber core as per (12). The rays traced using the gradient RI profile, shows bending of rays in the fiber core area. (Refer Supporting information Sec. F for more details.)

*3) Core RI profile for various bend ratios*

The effective RI of the optical fiber core at the outer and the inner curvature of a bent POF was calculated as per (12) for various bend ratios and is represented in Fig. 6. As anticipated, the effective RI of the core deviates from that of a straight fiber prominently (>2%) for bend ratios below 7. Hence, this regime (below ~7 bend ratio) where the plastic deformations are prominent is termed as the plastic regime. It is important to note that the effect of BIRI inhomogeneity on the light propagation, is also a function of bend ratio (refer equation 12 and Supporting information section F) and is negligible in the geometric (bend ratio >17) and even the saturation bending regime (>7). Thus, the results discussed in section III.A (for bend ratio >7) are not influenced by the inclusion of BIRI inhomogeneity effects.

Ray tracing by applying inhomogeneous core material properties (as shown in Fig. 5) is computationally more intensive than tracing rays in a homogeneous core. Hence, a simple approximation is used to improve computational efficiency. The tracing of rays is important to determine the angle at which the rays strike the outer curvature, which in turn determines whether the ray is guided or lost. However, it was found that instead of tracing the entire ray, the effective RI at the outer curvature could be substituted in place of $n_{core}$ in the Snell's law to calculate the critical angle condition and determine the ray guidance in a bent fiber. This approximation reduces the computational time drastically without compromising on the accuracy of the results (as verified by simulations).

*4) RI Sensitivity due to BIRI inhomogeneity*

The RI sensitivity of U-bent FOS incorporating the effects of bend-induced RI inhomogeneity is shown in Fig. 7. The simulation results show no considerable deviation in the RI sensitivity over the gentle ($R/\rho > 25$) and geometric ($17 < R/\rho < 25$) bend regimes from that of the model considering refractive losses due to geometry alone. However, in the plastic bend regime ($R/\rho < 7$), the increase in sensitivity was much more drastic in contrast to that observed in the geometric bend regime. It can be observed that the sensitivity reached a maximum value and any further decrease in the bend ratio resulted in significant loss of sensitivity. The point of maxima in the sensitivity curve determines the optimum bend ratio. This value is dependent on the RI of the surrounding aqueous medium. For example, in an aqueous medium of RI 1.36, the sensitivity of U-bent POF probes peaks for the bend ratio of ~1.69 and becomes zero at ~1.40. This value is also dependent on the material of the optical fiber (for silica optical fiber, the peak is at 3.4 and becomes zero at 2.5). Below the optimum bend ratio, the core RI tends to become lower than that of the surrounding medium, implying that all the light rays are lost, and the fiber cannot guide any meridional rays at the plane of bending. The RI sensitivity drops significantly as we move away from the optimum value. This signifies the importance of a suitable automated fabrication process for improved repeatability. The light propagation characteristics, the material properties, the resultant optical properties and the sensor parameters are significantly different for each regime. The constant bend ratio lines demarking these different regimes further establishes the importance of 'bend ratio' for sensing application. (Refer Supporting information Sec. G and H for more details.)

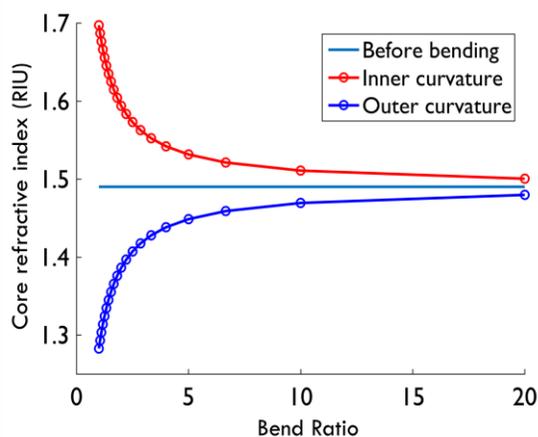

Fig. 6: The maximum and minimum of modified core refractive index due to bending at various bend radii

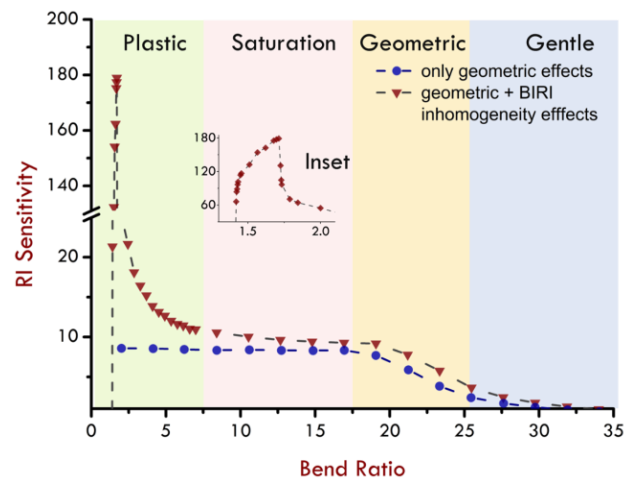

Fig. 7: RI Sensitivity of 0.5 mm POF U-bent probe, for varying bend ratios. Inset: RI Sensitivity for bend ratios between 1.4 and 2.1

Figures 4 and 7 show unrealistic values of RI sensitivity of U-bent POF probes. The absolute sensitivity values obtained by simulations are an overestimate. This could be due to several reasons including: (i) consideration of only 2D geometry and meridional rays as opposed to a more realistic consideration of 3D geometry and other variety of rays (ii) overestimation of BIRI inhomogeneity, since during the process of probe fabrication (heat treatment and bending), there will be movement of material such that the residual stress or inhomogeneity is minimized. Experimental evaluation of the material and photoplastic properties of the U-bent POF would be necessary for a more precise estimation of BIRI inhomogeneity and for better accuracy of the current model.

*C. Comparison with experimental data*

It is important for the model to be able to capture the qualitative trends observed in the experiments. This is essential to (i) validate the model, (ii) use the model to identify the optimum geometric and material parameters necessary to fabricate probes with maximum RI sensitivity.

Fig. 8 shows the comparison between the simulation results (using this model) and our experimental results (published elsewhere [5]). In comparison to Fig. 4 (obtained considering only geometry effects), the incorporation of BIRI inhomogeneity helps in understanding the high sensitivity exhibited by these probes at low bend ratios (Fig. 8D). Although the absolute sensitivity values predicted by the model significantly differ from that of the experimental data, the model is able to predict the trend of the probe behavior for various bend ratios. Fig. 8 A – C present a more detailed comparison of refractive losses for U-bent FOS with respect to (i) bend ratios (2.5, 3, 5, 7) and (ii) the fiber diameter (0.25, 0.5 and 0.75 mm). The results from simulations and experiments were plotted at different scales (however consistent for all three plots), mainly to compare the trends.

From Fig. 8A and B, it can be seen that the simulation result (scale invariance of bend ratio) corroborates with experimental results significantly. The deviations of the experimental results from the simulations could be due to the reasons stated in section III.B.4 and due to the manual fabrication processes through which experimental results were obtained.

IV. CONCLUSION

A ray optics based theoretical model was developed for understanding the nature of light propagation and the optical losses in a U-bent fiber optic sensor. A simple 2D geometrical representation of the U-bent FOS as a semi-circular ring was used to study the effect of meridional rays (propagating at the plane of bending) on the RI sensitivity of the sensor probes. This study differs from the previous ray optics models in the (i) representation of results with respect to bend ratio (independent of the particular fiber and bend radii), (ii) regimes of bending that are considered in the study, as most of the ray optics models developed earlier do not report their findings for bend ratios <5, (iii) detailed incorporation of material deformation effects.

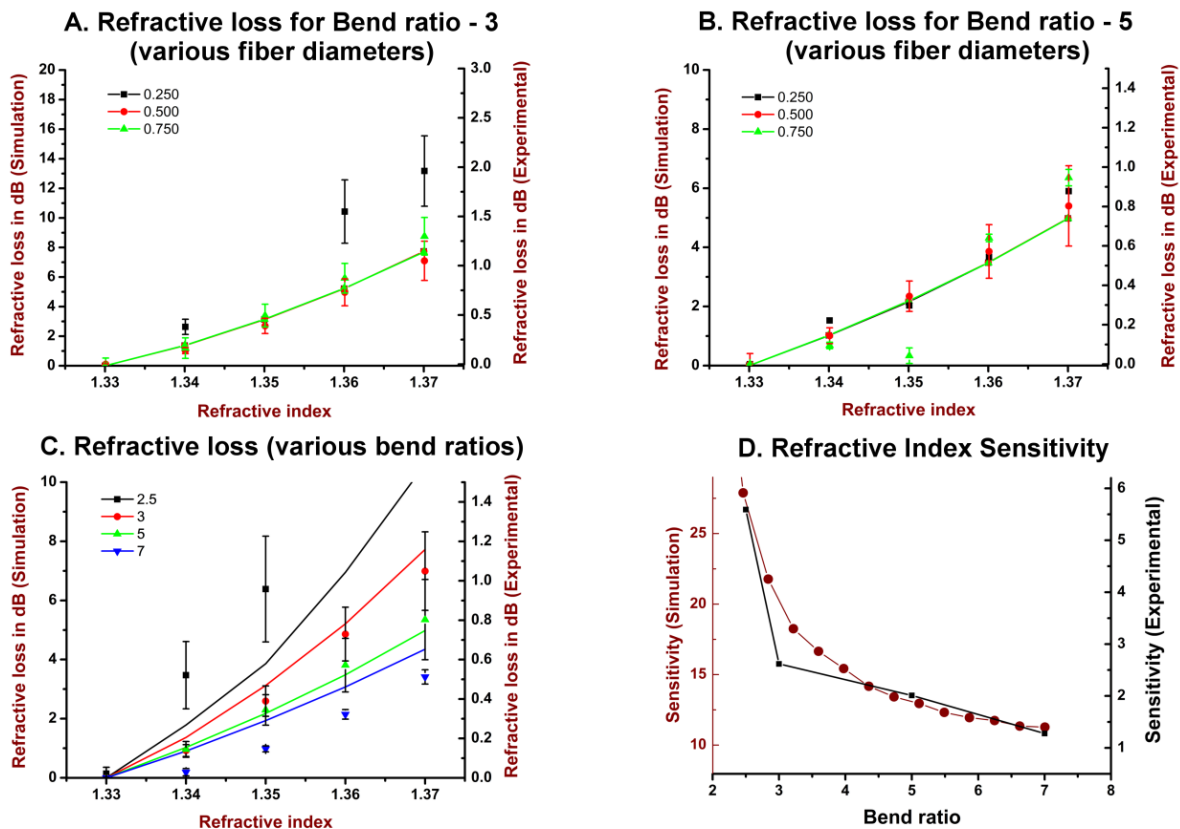

Fig. 8: Comparison between the results obtained through simulations (lines) with the experimental results (points) reported by Gowri and Sai 2016. **A and B**. Normalized refractive loss of U-bent FOS of various fiber diameters, for bend ratios 3 and 5 respectively. **C**. Normalized refractive loss of 0.5 mm U-bent FOS for various bend ratios. **D**. Sensitivity of U-bent 0.5 mm diameter FOS calculated at 1.37 RI. Experimental data was digitized from (Gowri and Sai 2016)

This study also establishes the importance of considering the mechanical and thermal stresses on the optical property of the bent fiber probes in contrast to just relying on geometric effects for explaining such high sensitivity. Bend Induced Refractive Index (BIRI) inhomogeneity in the optical fiber core resulting in lower RI at the outer interface of the fiber core is deduced to be the main contributing factor for high sensitivity of U-bent FOS, for bend ratios below 7. Optimum probe parameters are estimated for achieving maximum sensitivity, and the results obtained from the ray optics model are compared with experimental results and validated. Further, an important contribution of this study is the approach of simultaneously using both semi-analytical and numerical methods in understanding the underlying light propagation characteristics. The model could further be improved by considering 3D geometry and including skew rays in the study.

U-bent FOS have been widely used not just for RI sensing, but for evanescent wave-based absorbance sensing as well. In this regard, this U-bent FOS model could be extended beyond estimation of refractive losses to estimate power availability at the surface of the optical fiber probe.

# Supporting information for
# Investigating the refractive index sensitivity of U-bent fiber optic sensors using ray optics

Christina Grace Danny, M Danny Raj, VVR Sai

## A. Optical source and coupling

The incident power profile P(θ), is modeled using (1) and is shown in Fig. S1.

$$P(\theta) = \frac{n_1^2 \sin\theta \cos\theta}{(1 - n_1^2 \cos^2\theta)^2} \quad (1)$$

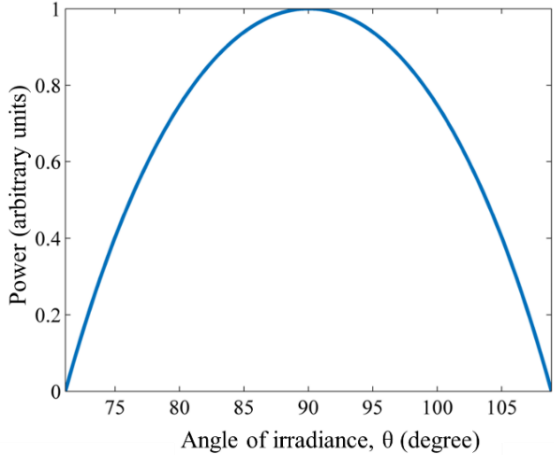

Fig. S1: Power distribution for rays coupled to an optical fiber using a microscopic objective

## B. Fresnel's transmission coefficients

Fresnel's coefficients are used in this ray tracing algorithm to precisely account for the refraction and reflection of the rays at the core-medium interface. The critical angle ($\theta_c$) that defines the conditions for total internal reflection is given by (2). Especially, when $\theta_o > \theta_{cc}$ a ray may neither be entirely refracted nor completely guided. The ray undergoes partial refraction and partial reflection at the core-medium interface. The partly reflected ray may undergo multiple reflections as it propagates further in the decladded bent region and loses almost all of its power. Fresnel's transmission coefficient (T) is used in the model to estimate the loss of light at each reflection as per (3). $P_{guided}$ in (4) refers to the partial power of a reflected ray that reaches the other end of the U-bent FOS. When $\theta_o < \theta_{cc}$ the ray is guided by the optical fiber and no transmission losses are considered for such a case.

$$\sin\theta_c = \left(\frac{n_{med}}{n_{co}}\right) \quad (2)$$

$$T = \frac{4\sin\theta_o}{\sin\theta_c} \sqrt{\left[\frac{\sin^2\theta_o}{\sin^2\theta_c} - 1\right]} \quad (3)$$

$$P_{guided}(\theta_{refracted}) = P(\theta) \times \exp(-T) \quad (4)$$

## C. Convergence of model parameters

The accuracy of a ray tracing model is highly influenced by the number of rays considered in the simulation, which is a product of $N_a$ and $N_p$ (where $N_a$ is the number of angles considered at each point source and $N_p$ is the number of point sources distributed over the cross-section). Ideally, the number of rays (based on V number) is calculated as per (5).

$$N = 0.5 \times \left(\frac{\pi \times Fiber\ Diameter \times Numerical\ aperture}{\lambda}\right)^2 \quad (5)$$

For typical values of wavelength (λ), fiber diameter and numerical aperture (0.6 µm, 0.5 mm, 0.48), the number of rays correspond to $6.7 \times 10^5$. However, simulations involving such a large number of rays and optimization algorithms to identify optimum probe parameters are computationally intensive. Hence, the minimum number of rays sufficient to simulate the optical coupling is determined by examining the convergence of output parameter for a range of rays ($N_a \times N_p$) from 100 to 40000. The output parameter (absorbance) converges below a value of $N_a = 100$ and $N_p = 100$ as shown in Fig. S2. Hence, all the subsequent simulations were carried out with these values for $N_a$ and $N_p$, implying 10,000 rays per simulation.

Christina Grace Danny was with the Department of Applied Mechanics, Indian Institute of Technology Madras, Chennai 600036, India and is currently with the Department of Electronics and Instrumentation, Ramaiah Institute of Technology, Bengaluru 560094, India. (e-mail: christina@msrit.edu).

M Danny Raj was with the Department of Chemical Engineering, Indian Institute of Technology Madras, Chennai 600036, India and is currently with the Department of Chemical Engineering, Indian Institute of Science, Bengaluru 560094, India (e-mail: dannyrajmasila@gmail.com, dannym@iisc.ac.in).

V V Raghavendra Sai is with the Department of Applied Mechanics, Indian Institute of Technology Madras, Chennai 600036, India (e-mail: vvrsai@iitm.ac.in).

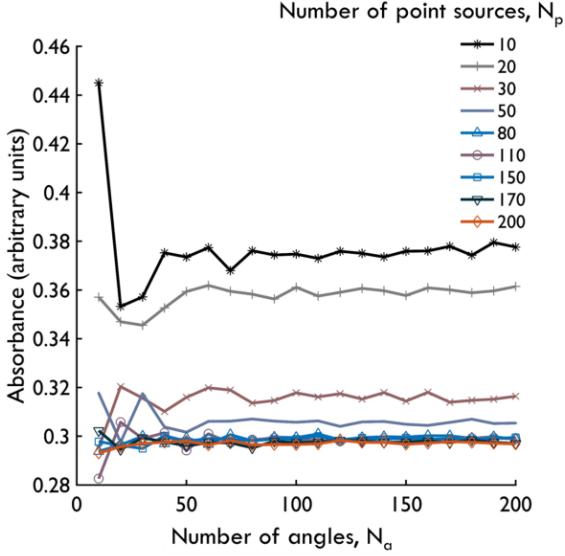

Fig. S2: Absorbance simulated using the 2D ray tracing model for a 500 μm POF U-bent probe with a bend diameter of 3.0 mm and aqueous medium RI of 1.36. Simulations have been carried out for varying number of ray positions and angles of propagation to verify convergence of resultant absorbance

### D. Analytical and numerical estimation of local NA

The analytical expression for local numerical aperture (NA), that is characteristic of the light guidance capacity is given as per (6).

$$\mathrm{NA} = \sqrt{\left[n_{co}^2 - n_{med}^2 \left[\frac{R+\rho}{R+x}\right]^2\right]} \quad (6)$$

The distance at which the NA becomes 0 as estimated by the analytical equation (6) significantly matched with the width of propagation as evaluated from the numerical 2D ray tracing algorithm. The NA across the various bend radii for RI of 1.44, (for RI = 1.37, is given in the main article) calculated using analytical equation and compared with the width of propagation estimated using ray optics model is shown in Fig. S3.

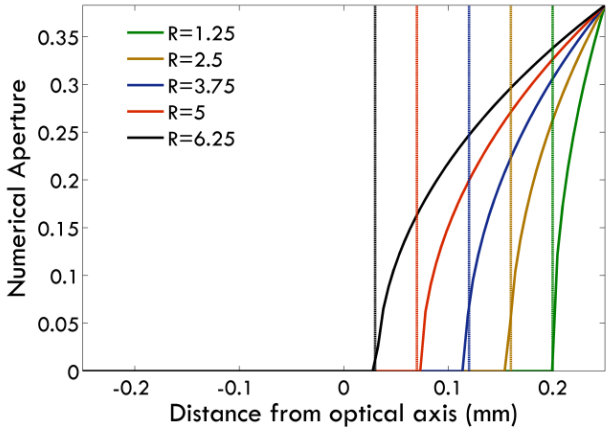

Fig. S3: Simulations results obtained for a plastic optical fiber of 0.5 mm core diameter. The NA across the various bend diameters for RI of 1.44, calculated using analytical equation and compared with the width of propagation estimated using ray optics model

### E. Estimation of transition radius

In this section, the transition radii, $R_{geometric}$ and $R_{saturation}$ that demark the geometric bending region (Fig. S4) are additionally estimated by two methods other than the numerical ray tracing method discussed above. For analytical estimation considering only geometric effects, we begin by noting that the RI sensitivity is determined by the number of rays that are lost due to refraction. As mentioned earlier, it can be seen from the ray optics model that all the rays that are lost in the bend region are lost at the outer interface. Thus, the extent of refractive loss may be evaluated by estimating the modified angle at the outer interface, $\theta_o$, using a simple trigonometric equation (7) reported by Gupta et al [17].

$$90 - \theta_o = \sin^{-1}\left[\left(\frac{R+x}{R+\rho}\right)\sin\theta\right] \quad (7)$$

From (7), it can be seen that the modified $\theta_o$ is a function of (i) x, the location of the point with reference to the optical axis (as given in Fig. 1), (ii) $\theta$, the irradiance angle, (iii) $\rho$, the fiber radius and (iv) R, the bend radius. In a straight optical fiber, (1) describes the power distribution for the range of irradiant angles (as shown in Fig. S4). In a bent optical fiber, (7) can be used to estimate the modified angle at the bend region, while the power distribution can be calculated by weighting it with the power profile given in (1) (as shown in Fig. S4). The rays with modified angle $\theta_{o, mode}$ are the rays that are most frequent, carrying the maximum optical power. $\theta_{o, mode}$ for various bend ratios is estimated and the complimentary angle ($\theta_{oc, mode}$) is represented in Fig. S4. It can be seen that the ($\theta_{oc, mode}$) curve intersects the straight line representing the critical angles ($\theta_{o,guided}$) for $n_{med}$=1.33 (water, as reference) and 1.37 at ~24.3 and ~17.6 bend ratios respectively. From Fig.S4, it can be seen that, these values match the $R_{geometric}/\rho$ and $R_{saturation}/\rho$, as achieved through the ray tracing algorithm.

From the above estimation, it could be interpreted that: (i) for bend ratios above 24, the alteration of the angle due to bending is minimal, such that ($\theta_{oc, mode}$) is greater than the minimum guidance conditions required for $n_{med}$= 1.37 resulting in negligible refractive losses. Thus, the RI sensitivity in this regime is very low; (ii) for the bend ratios below 24, the modified angle ($\theta_{oc, mode}$) is lower than the guidance requirement of rays in $n_{med}$= 1.37 but higher than that in water (RI = 1.33), resulting in significant medium-sensitive refractive losses; (iii) for bend ratios below the transition radius $R_{saturation}/\rho$, the modified angle ($\theta_{oc, mode}$) is lower than the guidance requirement for the reference medium. Thus, bending below the transition radius results in a significant loss of rays even in the presence of the reference medium, allowing for only a fractional rise in the medium-sensitive refractive loss, expressed as plateau in the sensitivity curve. The estimation of the extent of modification in the angle of propagation could thus explain the trend observed in Fig. S4.

Alternatively, as numerical aperture depicts the light carrying capacity of the entire optical fiber, analysis was undertaken pertaining to the NA of bent fibers. From Fig. S3, it can be seen that an optical fiber bent at R = 15 behaves as a step index (two boundary) waveguide, while a fiber bent at R = 9 behaves like a single boundary waveguide. The transition ratio at which the bent optical fiber transits into a single boundary waveguide was estimated by equating the local NA at the inner curvature to zero in (8).

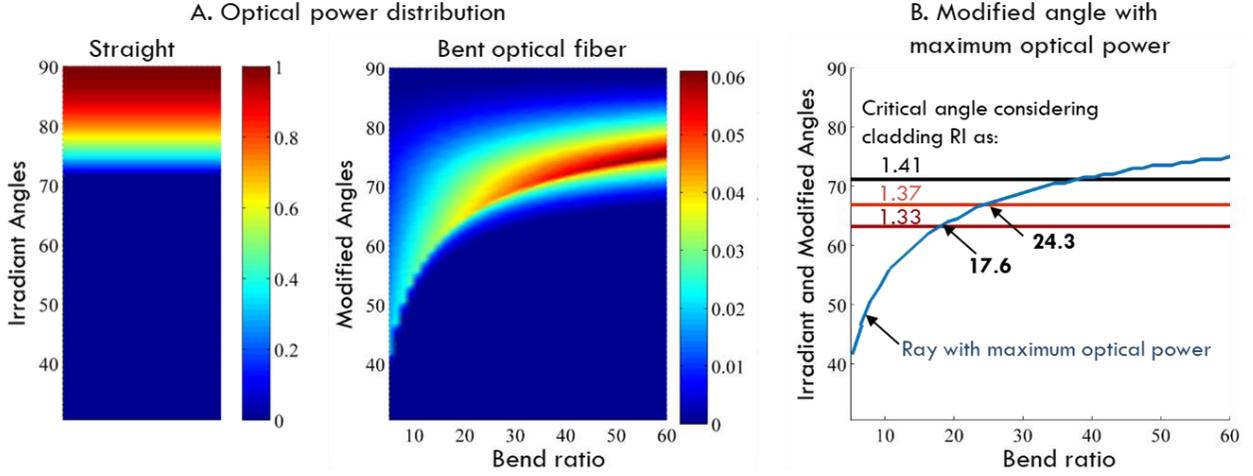

Fig. S4: A. Optical power distribution across various irradiant angles (straight region), modified angles (bent region) for different bend ratio, B. Estimation of transition radius based on the modified angle $\theta_{oc,mode}$ that is most frequent (consequently carrying maximum optical power) plotted for various bend ratios. The intersection of this curve with the straight lines denoting critical angles of U-bent FOS in water and 1.37 RI are 24.3 and 17.6 respectively.

$$\mathrm{NA} = \sqrt{\left[n_{co}^2 - n_{ref}^2 \left[\frac{R+\rho}{R-\rho}\right]^2\right]} = 0 \quad (8)$$

$$\left[\frac{R+\rho}{R-\rho}\right] = \left[\frac{n_{co}}{n_{ref}}\right] \quad (9)$$

$$\left[\frac{R}{\rho}\right] = \left[\frac{n_{co}+n_{ref}}{n_{co}-n_{ref}}\right] \quad (10)$$

For conditions mentioned in section III.A.2 and III.A.3, the bend ratio for which NA transits to single boundary waveguide for $n_{ref}$ = 1.33 and $n_{ri}$ = 1.37 is estimated to be 17.6 and ~24.8 respectively. This result substantiates that geometric losses are prominent only in the region between saturation and geometric bend radii. Both the approaches mentioned here (estimation of $\theta_{o-mode}$ and local NA) estimate transition radii $R_{geometric}$ and $R_{saturation}$ as ~25 and ~17 respectively, validating the result obtained using numerical ray tracing.

### F. Bending of light ray due to inhomogeneous core

For incorporation of BIRI inhomogeneity, the ray tracing algorithm was modified to accommodate the gradient RI profile of the optical fiber core as per (8).

$$n_x = 1.49 \left(1 - 0.1391 \times \left[\frac{x}{R}\right]\right) \quad (11)$$

Ray propagation was traced in minute steps (of 0.01 times the fiber diameter), evaluating the angle change and propagation direction at each step, using the estimated RI of the fiber core at each spatial point. Fig. S5 shows rays traced using the gradient RI profile, showing bending of light rays in the fiber core area.

It may be noted that there is a significant difference in the path traced by a ray in a bent optical fiber versus a graded index fiber. This is due to (i) the unidirectional gradation due to BIRI inhomogeneity (increase in RI at the inner curvature and decrease in RI at the outer curvature) and (ii) the geometrical shape of the fiber resulting in such a ray path as described in Fig. S5. The curved (non-linear) trajectory of the light rays was prominent for acute fiber bends and insignificant for larger bend ratios.

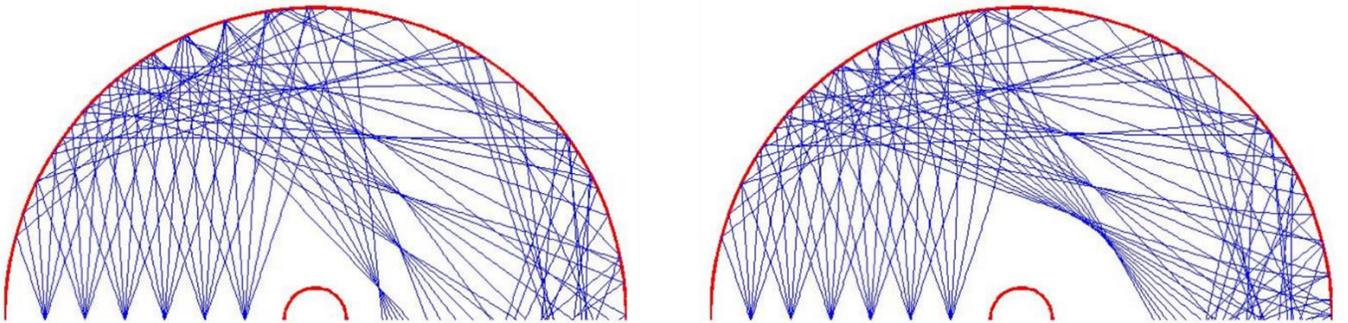

Fig. S5: Linear (left) and curvilinear (right) propagation of the light rays in the fiber core with homogeneous and bend-induced inhomogeneous refractive index profile respectively

### G. Dependence of RI sensitivity on the Bend Ratio

The ray optics model developed here uses a simple 2D geometrical representation of the U-bent FOS as a semi-circular ring. The optical source is modeled as a large number of point sources distributed over the cross-section of the fiber. Thus, considering the geometry of the system, bend ratio (the ratio of bend radius to fiber radius) is a dimensionless quantity that makes the system scale invariant.

The BIRI inhomogeneity effects are also a function of the ratio of distance from the center of the optical fiber to the bend radius (8). Thus, inclusion of BIRI inhomogeneity doesn't change the scale invariance of the system. RI sensitivity of U-bent FOS with various Fig. S6 shows the value of RI sensitivity for various combinations of bend radii and fiber radii (in a 2D scatter plot with RI sensitivity values in color). It can be observed that the RI sensitivity values remain the same for constant ratios of bend radius to fiber radius (shown in Fig. S6 as dashed lines with scatter values of same color). The RI sensitivity value increases as the bend ratio decreases, as elaborated in section III.B.4 and III.D.

### H. Bending regimes

The bending of optical fibers in the field of communication applications are usually characterized as macro-bends or micro-bends [37]. Here, macro-bends are a resultant of macroscopic coiling of cables (>15-20$\rho$) and micro-bends refer to deformation of the circular cross section of the optical fiber due to external pressure. However, in case of deliberate and controlled bending of optical fiber for sensing application, we propose the following bending regimes based on the behavior of the probes. Based on the simulation results, the parametric space of bend radius to fiber radius is divided into four regimes namely (i) gentle (ii) geometric (iii) saturation and (iv) plastic (Fig. S7). As the fibers are bent and the bend ratio reduces to 25, the RI sensitivity is mostly unaffected. This region is named *gentle*. However, a further reduction down to 17 considerably improves the sensitivity due to the refractive losses signifying the geometric effects. This region is named *geometric*. Further bending has minimal influence of geometric effects on the RI sensitivity. Hence the region is named *saturation*. For the bend ratios below 7 down to 1, the fiber core experiences a significant material deformation leading to core refractive index changes and results in an exponential increase followed by an abrupt drop in RI sensitivity. This region is named as *plastic*. The minimum possible bend radius equals the fiber radius, thus the region where R is less than $\rho$ is marked as *impossible*. Estimation of the radii of demarcation are elaborated in the section III.A.3.

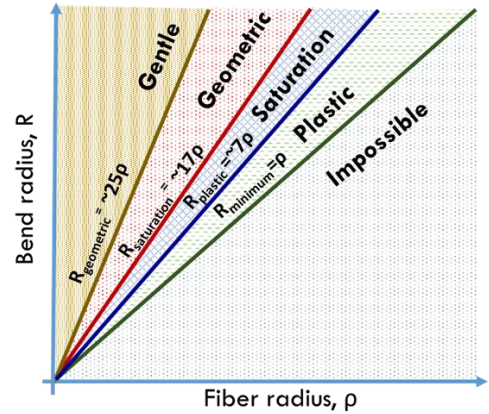

Fig. S7: Parametric space of bend radius and fiber radius showing the different regimes of bending. The solid lines represent constant ratios of bend radius to fiber radius.

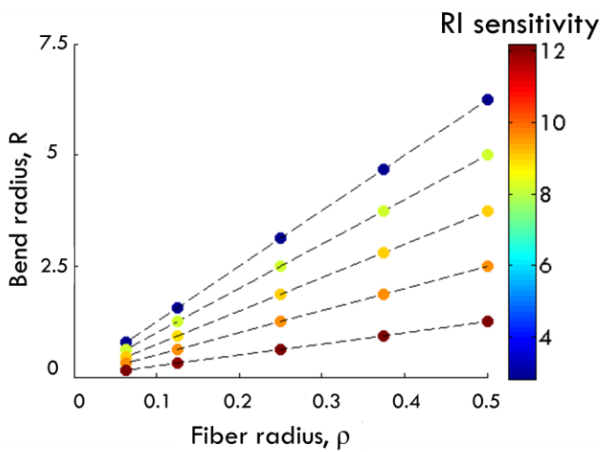

Fig. S6: 2D scatter plot with RI sensitivity values in color, obtained for a plastic optical fiber of bend radii and fiber radii in the range of (0.3125 – 12.5) and (0.125 to 1.00 mm) respectively. The dashed lines represent constant bend ratios (ratio of bend radius to fiber radius).

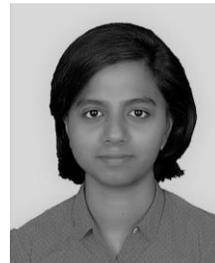

**Christina Grace Danny** received her B.Tech. in Instrumentation and Control from the National Institute of Technology Tiruchirappalli, Tamil Nadu, India in 2009 and M.S. and Ph.D. degree in Applied Mechanics from Indian Institute of Technology Madras, Chennai, India in 2018.

From 2009 to 2011, she was working as an Instrumentation Engineer in Mangalore Refineries and Petrochemicals Limited, Mangalore, Karnataka, India. Since 2018, she is an Assistant Professor in Ramaiah Institute of Technology, Bengaluru, Karnataka, India. She has authored 2 articles and 4 conference proceedings. Her research interests include optical fiber sensing, thin film coatings, plasmonics, geometric optics modeling and SERS.

She received a best poster presentation award and a trophy for excellence in creativity and collaboration at IC-IMPACTS Summer Institute on Optical Solutions for water, infrastructure and health (June 2015), held at University of Toronto, Canada.

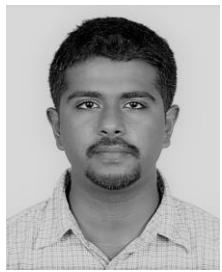

**Danny Raj M** received his PhD from the Indian Institute of Technology Madras, India (IIT-M) in 2017 for his work on understanding collective droplet phenomena in microchannel systems using agent-based models. After his PhD, he continued in IIT-M as a post-doctoral fellow, setting up experiments for high throughput particle synthesis and developing mathematical models for industrial processes (in collaboration with GE). His research interests include, modelling and simulation of the dynamics of physicochemical systems and collective phenomena in natural and artificial systems.

In 2017, he was awarded the INSPIRE faculty award (by Department of Science and Technology)- which is 'Assured Opportunity for Research Career (AORC)' for young researchers. He is currently hosted in the department of chemical engineering at the Indian Institute of Science Bangalore, India as an INSPIRE faculty where he pursues research in a wide variety of topics that include *granular matter*- understanding the motion of an intruder through a crowded environment *traffic flow*- motion of agents with cognition and *ecology*- identifying underlying interactions in a school of fish (with CES, IISc). He has authored 6 peer reviewed articles and 1 conference proceeding.

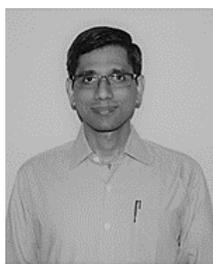

**V.V.R. Sai** received his Ph.D. in Biomedical Engineering from Indian Institute of Technology, Bombay, India in 2009 for the work done on development of two novel fiber optic biosensing technologies for detection of biomolecules and pathogens.

He spent 2 years of postdoctoral fellowship at University of Idaho, USA working on development of SERS based DNA biosensor, receptor mediated detection of explosives and application of nanomaterials for development of drug delivery systems for anti-sense and anti-gene oligonucleotide based therapeutics.

He came back to India and joined IIT Madras in August 2011 as an Assistant Professor of Biomedical Engineering in Department of Applied Mechanics at IIT Madras, where he established biosensors laboratory to focus on development of affordable and indigenous technologies for clinical diagnostics, water, food and environmental monitoring. Since July 2017, he is working as an Associate Professor.

He has authored 22 peer reviewed research articles, more than 25 conference presentations/proceedings and filed 3 patents. He is also a recipient of prestigious Young Engineer Award from Indian National Academy of Engineering (INAE) and travel award from Epson Research Foundation in 2015 and 2016 respectively. Currently, a postdoc and five Ph.D. research scholars are working with him.